\documentclass[12pt]{article}
\usepackage{subcaption}
\providecommand{\keywords}[1]{\textbf{Keywords:} #1}
\usepackage{geometry}
\usepackage[utf8]{inputenc}
\usepackage{amsfonts}
\usepackage{latexsym}
\usepackage{amssymb,latexsym}
\usepackage{amsthm}
\usepackage{indentfirst}
\usepackage{graphicx}

\usepackage{fancyhdr}
\usepackage{hyperref}
\usepackage{cite}
\usepackage{amsmath}    
\usepackage{amsfonts}   
\usepackage{amssymb}    
\usepackage{graphicx}   
\usepackage{geometry}   
\usepackage{hyperref}   
\usepackage{amsthm}
\makeatletter
\renewcommand{\thefigure}{\textbf{\@arabic\c@figure}}
\makeatother
\makeatletter
\renewcommand{\thetable}{\textbf{\@arabic\c@table}}
\makeatother
\makeatletter
\def\hlinewd#1{%
\noalign{\ifnum0=`}\fi\hrule \@height #1 %
\futurelet\reserved@a\@xhline} \makeatother

\newcommand{\set}[1]{\left\{\right\}}

\newcommand{\re}[1]{(\ref{})}
\newcommand{\nd}{\noindent}

\newcommand{\tn}[1]{\textnormal{}}

\newtheorem{lem}{Lemma}[section]

\newtheorem{exm}{Example}[section]

\begin{document}
\title{A New Regression Model for Analyzing Non-Stationary Extremes in Response and Covariate Variables with an Application in Meteorology}
\author{Amina EL BERNOUSSI and Mohamed EL ARROUCHI \\ Ibn Tofail University, Faculty of Sciences Department of Mathematics, Kenitra, Morocco\\
E-mail: amina.elbernoussi@uit.ac.ma}

\date{}
\maketitle
\begin{abstract}
\nd The paper introduces a new regression model designed for situations where both the response and covariates are non-stationary extremes. This method is specifically designed for
situations where both the response variable and covariates are represented as block maxima, as the limiting distribution of suitably standardized componentwise maxima follows an extreme value copula. The framework focuses on the regression manifold, which consists of a collection of regression lines aligned with the asymptotic result. A Logistic-normal prior is applied to the space of spectral densities to gain insights into the model based on the data, resulting in an induced prior on the regression manifolds. Numerical studies demonstrate the effectiveness of the proposed method, and an analysis of real meteorological data provides intriguing insights into the relationships between extreme losses in precipitation and temperature.

\end{abstract}
\nd \keywords{Spectral measure. Bivariate extreme value distribution. Joint distribution. Quantile regression. Non-stationary extremes}
\section{Introduction}
\nd In recent years, there has been much interest in statistical techniques for simulating extremes using non-stationary time series—that is, series with statistical characteristics that change over time due to dynamic system changes. It is more beneficial in various applied disciplines, particularly in methodology related to earth science \cite{Ch14,Ka13,Men16}. We should be cautious of the harm and disasters resulting from climate extreme occurrences that impact society and the economy. Because both are dynamic, the non-stationary time series can be used to describe the fluctuations in climate.\\
We have fitted a new model in this research that explicitly develops a regression procedure for the case where both the response and the variables are severe and non-stationary. It has to do with quantile regression, which was initially created in 1978 by Koenker and Bassett. The original goal of the technique was to model the conditional quantile of a response $Y$ given a covariate $X=(X_1,..., X_p)^T$, that is, where
\begin{equation}\label{eq1}
F^{-1}(q|x)=\inf\{y: F(y|x) \geq q\},\;\;\;\;\;\; 0<q<1,
\end{equation}
and $F(y|x)$ is the distribution function of $Y|X=x.$ We began with a significant finding in extreme value statistics, which shows that the appropriately standardized vector of block maxima converges in distribution to an extreme value copula introduced by Gudendorf and Segers in 2010. Regression manifolds involve a group of regression lines that illustrate the relationship between extreme values. It helps to understand how an extreme value covariate can impact an extreme value response, as explained by de Carvalho et al. \cite{Ca22}. We have also proposed a prior in the regression manifold space by resorting to a Logistic-Normal distribution on the space of spectral densities proposed by Aitchison et al. \cite{ait80}. The primary distinction is that our regression model is designed for scenarios where both the response and covariates are non-stationary extremes. Its objective is to understand how the limiting distribution of a suitably standardized block maxima response  $(M_{x})$ is influenced by a covariate  $\mathbf{x} =(x_1,...,x_p)^T,$ through the specified model
\begin{equation}\label{eq2}
(M|X=x)\sim GEV(\mu_x,\sigma_x,\xi_x),
\end{equation}

\nd i.e., the parameters location, shape, and scale are not constants; they depend on the time \cite{Ka13,Col01,Ea09,Ka02,St05,Wa09}. A fascinating avenue for exploration would involve investigating non-stationary techniques about other methodologies, such as those outlined by Goda \cite{Go88} and Boccotti \cite{bo00}, which are grounded in the transformed-stationary approach established by Menstaschi et al. \cite{Men16}, which was adopted for non-stationary extreme analysis. Our approach also works with situations where both the response and covariates are stationary extreme (i.e. the location, scale, and shape parameters $\mu\in \mathbb{R}$, $\sigma >0$, and $\xi\in \mathbb{R}$ respectively) but in this paper we are focused on the non-stationary case for modeling the climatic extremes later. Our current work is an alternative to the work of Carvalho et al. \cite{Ca22} in the non-stationary case. The main difference is that the Logistic-normal distribution is used instead of the Dirichlet distribution. The Logistic-Normal distribution is a valuable tool for analyzing temperature and precipitation data in environmental research. Its flexibility, ability to model dependencies and support for Bayesian analysis make it suitable for understanding complex relationships in climatic data, ultimately aiding in better decision-making and predictions in environmental management and policy. Aitchison \cite{ait80} proved that it offers a richer parameterization than the Dirichlet distribution, which is often used for compositional data. This flexibility allows researchers to capture complex relationships and correlations between temperature and precipitation data, which may not be independent. Also, it facilitates straightforward estimation and hypothesis testing for environmental data. Moreover, it supports Bayesian methods, allowing for the incorporation of prior knowledge about temperature and precipitation patterns. This is particularly useful in predictive modeling and assessing future climate scenarios. While the Dirichlet distribution is commonly used for modeling proportions, it assumes independence among components, which may not be realistic for temperature and precipitation data. With its ability to model dependencies, the Logistic-Normal distribution provides a more accurate representation of the underlying processes.

\nd To demonstrate the effectiveness of our model, we conducted both qualitative and quantitative comparisons with the model proposed by Carvalho et al. \cite{Ca22}. To achieve this, we utilize the criteria Akaike Information Criterion (AIC) introduced by Akaike \cite{ak74} and the Bayesian Information Criterion (BIC) proposed by Schwarz \cite{SC78}. Fortunately, these two famous criteria validate our model's effectiveness and goodness. Ultimately, we implemented our novel approach using actual data from methodology from local Meknes-Morocco, where we considered extreme precipitation as a response and extreme temperature as covariates.

\nd Whereas our paper is divided as follows. In Sect.\ref {sec2}, we look at the background of the non-stationary GEV model, the famous transformed-stationary approach, and some properties of the Logistic-Normal distribution. Also, we present a background on regression manifolds and some parametric instances. Then, we introduce the proposed model, which induces a new prior on the space of regression manifolds. Moreover,  in Sect.\ref{sec3} we simulate that with real data of meteorology, and we consider the precipitations as a response and the temperature as a covariate, and summarize with a little conclusion for our new approach.  \\

\section{Modeling conditional bivariate extreme value distribution}\label{sec2}
\subsection{Background on non-stationary GEV model and the transformed-stationary approach}\label{sec2.1}
\nd Extreme Value Theory (EVT) offers a robust framework for analyzing climate extremes and determining their corresponding return levels \cite{Col01}. EVT posits that, under certain conditions, the distribution of extreme values—encompassing both maxima and minima—approaches one of three limiting distributions: Gumbel, Fréchet, or Weibull. Together, these distributions comprise the Generalized Extreme Value (GEV) distribution. Various studies have utilized GEV to explore extreme events \cite{Ka13}. This methodology is often called the block maxima approach. An alternative method within EVT called the peak-over-threshold (POT) approach, centers on assessing extreme values that surpass a defined high threshold, employing a generalized Pareto distribution \cite{Sm87}. Both the annual maxima and POT techniques are extensively applied in the study of extreme climate phenomena \cite{Vi11}.

\nd The cumulative distribution function of the GEV can be represented as follows

\begin{equation}\label{eq3}
F(x;\mu,\sigma,\xi)=\exp\left\{-\left(1+\xi\left(\frac{x-\mu}{\sigma}\right)\right)^{-1/\xi}\right\},\;\;\;\left(1+\xi\left(\frac{x-\mu}{\sigma}\right)\right)>0.
\end{equation}

\nd The Generalized Extreme Value (GEV) distribution is a versatile model for capturing extreme behaviors and is defined by three parameters: $\mu$, $\sigma$, and $\xi.$ The location parameter $\mu$ indicates the center of the distribution; the scale parameter $\sigma$ measures the extent of deviations from $\mu$ while the shape parameter $\xi$ affects the distribution's tail characteristics. Specifically, as $\xi$ approaches 0, the distribution converges to the Gumbel distribution; when $\xi<0$, it corresponds to the Weibull distribution, and for $\xi>0$, it represents the Fréchet distribution.
Extensive research has been conducted on extreme value theory, particularly concerning stationary random sequences \cite{papa13}. In this context, stationarity is defined by the consistency of extreme properties over time \cite{Le83}. However, in non-stationary processes, the parameters of the underlying distribution become time-dependent \cite{ren13}, leading to varying distribution characteristics over time \cite{Me00}.\\

\nd For analyzing non-stationary extremes, we adopt a transformed-stationary approach (TS) recently developed by Mentaschi et al. \cite{Men16}, which estimates the parameters $\mu$, $\sigma$, and $\xi.$ The authors validate the TS methodology using various time series, including significant wave height and river discharge, demonstrating its effectiveness compared to established methods. This approach offers advantages such as utilizing the entire time series for estimation, simplifying the analysis by decoupling non-stationarity detection from distribution fitting and being easier to implement. This well-known methodology consists of three steps: first, transforming the non-stationary time series $y_t$ into a stationary series $x_t$; next, performing a stationary extreme value analysis (EVA); and finally, reverting the resulting extreme value distribution to a time-dependent form. The transformation $y_t \rightarrow x_t$ we propose is
\begin{equation}\label{1}
x_t=\frac{y_t-T_{y_t}}{C_{y_t}},
\end{equation}

\nd where $T_{y_t}$ represents the trend of the series, which is a curve indicating the long-term, slowly changing pattern of the series, and $C_{y_t}$ represents the long-term, changing the magnitude of a confidence interval that illustrates the distribution's magnitude of $y_t$. Specifically, if $C_{y_t}$ is equal to the long-term changing standard deviation $S_{y_t}$ of the series $y_t$, equation \eqref{1} simplifies to a straightforward time-varying adjustment of the signal
\begin{equation}\label{2}
x_t=\frac{y_t-T_{y_t}}{S_{y_t}},
\end{equation}

\nd with $T_{y_t} = T_{0y_t} + s_{T_t}$ and $ S_{y_t} = S_{0y_t} . s_{S_t}.$ In this context, $T_{0y_t}$ and $s_{T_t}$ refer to the long-term and seasonal variations of the trend, respectively. Meanwhile, $S_{0y_t}$ represents the long-term standard deviation, while $s_{S_t}$ indicates the seasonal factor of the standard deviation. So, we obtain
\begin{equation}\label{3}
x_t=\frac{y_t-T_{0y_t} - s_{T_t}}{S_{0y_t} . s_{S_t}}.
\end{equation}
The parameters of the time-varying GEV can be represented as
\begin{equation}\label{a}
\xi_y = \xi_x = \mathrm{const}.,
\end{equation}
\begin{equation}\label{b}
\sigma_{y_t} = S_{0y_t}.s_{S_t} . \sigma_x,
\end{equation}
and
\begin{equation}\label{c}
\mu_{y_t}  = S_{0y_t} . s_{S_t} . \mu_x + T_{0y_t}  + s_{T_t}.
\end{equation}
We can estimate a signal's slowly varying trend and standard deviation along with its seasonality using various methods. This proposal will use a simple approach based on a running mean and standard deviation. The trend $T_{0y_t}$ will be represented as a running mean of the signal  $y_t$ over a multi-year time window  $w$,
\begin{equation}\label{4}
T_{0y_t}  =  \sum_{tt =t -w /2 }^{tt =t +w /2} y_{tt}/N_t,
\end{equation}
where $N_t$ represent the number of observations available within the time interval $[t - w/2, t + w/2]$. To estimate the seasonality of the trend for a specific month of the year, we calculate the average monthly deviation from the detrended series. Specifically, for each month, the seasonality is determined by taking the average difference between the detrended series and the corresponding values
\begin{equation}\label{5}
s_T(month[t]) = \sum_ {years} \frac{[y_{tt} - T_{0y_{tt}}] |tt\in month(t) }{N_{month} },
\end{equation}

\nd where the subscript $tt \in the\; month[t]$ signifies that the averaging process is restricted to time intervals within each specific month of the year. For instance, the seasonality for January is calculated as the average of the detrended signal specifically for January. To determine the slowly varying standard deviation, we apply a running standard deviation using the same time window that was utilized to compute $T_{0y}(t)$:
\begin{equation}\label{6}
S_{0y_{t}}|ROUGH =  \sum_{t t =t -w /2}^{tt =t +w /2} \sqrt{[y_{tt} - \bar{y}_{tt \in[t - w/2, t + w/2]}]^2/N_{w_{sn}}},
\end{equation}
where the subscript "ROUGH" highlights the sensitivity of this expression to outliers, which can result in significant statistical errors when applied directly. To address this issue, we apply a moving average to smooth the equation \eqref{6} over a time window smaller than  $w$, such as $w/l$ where $l = 2$:
\begin{equation}\label{7}
S_{0y_{t}}  =  \sum_{t t =t -w /2l }^{t t =t +w /2l} LS_{0y_{tt}} |ROUGH/N_t.
\end{equation}
It is crucial to emphasize that further smoothing of the results from running means and standard deviations is advisable if it minimizes errors and enhances the detection of gradually evolving statistical patterns in the time series. This approach is necessary because estimating $T_{0y}(t)$ and $S_{0y}(t)$ involves employing a lowpass filter to smooth the signal over timescales shorter than $w$, effectively eliminating high-frequency fluctuations. To assess seasonality, we apply another running standard deviation, denoted as $S_{sn_{t}} $, using a time window  $W_{sn}$ that is significantly shorter than one year, typically around one month:
\begin{equation}\label{8}
S_{sn_{t}}=\sum_{t t =t -W_{sn} /2}^{tt =t +W_{sn} /2} \sqrt{[y_{tt} - \bar{y}_{tt \in[t - W_{sn}/2, t + W_{sn}/2]}]^2/N_t}.
\end{equation}
The seasonality of the standard deviation can be calculated as the monthly average of the ratio between $S_{sn_t}$ and $S_{0y_t} $
\begin{equation}\label{9}
s_S(month[t]) = \sum_{years} \frac{[S_{sn_{tt}}/S_{0y_{tt}}] |tt\in month(t)}{N_{tt} \in month(t )}.
\end{equation}
The estimated seasonality terms $s_T$ and $s_S$ are periodic with one year. We applied the equations \eqref{4}, \eqref{5}, \eqref{6}, \eqref{7}, \eqref{8}, and \eqref{9} in the equation \eqref{3}  for converting the non-stationary time series to a stationary series. The last step is that the fitted GEV extreme value distribution must be back transformed into a non-stationary one as given in equations  \eqref{a}, \eqref{b}, and \eqref{c}. For more details, see Mentaschi et al. \cite{Men16}.\\

\nd Now, we need to lay the groundwork on bivariate non-stationary extremes. In the statistical modeling of extreme value dependence, the spectral measure of a bivariate extreme value distribution is crucial. This paper introduces a model for the spectral measure that applies to any number of dimensions and allows for a generalization that includes mass on the simplex boundaries. To establish the foundation, let $\{(X_{it},Y_{it}\}_{i=1}^{n}$ with $t \in [i-\frac{n}{2},i+\frac{n}{2}]$
be a sequence of independent identically distributed random vectors with unit Fréchet marginal distributions, i.e. $\exp(-1/s_t)$, for $s_t>0$. In our case, $Y_{it}$ is considered a response, but $X_{it}$ is considered a covariate. Let the componentwise block maxima be $M_{nt}=(M_{x_t,n},M_{y_t,n})$ with $M_{y_t,n}=\displaystyle \max_{i=1,...,n}\{Y_{it}\}$ and $M_{x_{t},n}=\displaystyle \max_{i=1,...,n}\{X_{it}\}.$ According to the componentwise maxima method proposed by Coles et al. \cite{Col01}, the representation of the bivariate model becomes simple by assuming both margins have the
standard Fréchet distribution. The vector of normalized componentwise maxima $M_{nt}/n$ converges in distribution to a random vector $(X_t, Y_t)$, which adheres to a bivariate extreme value distribution characterized by a joint distribution function $G$. This distribution is non-degenerate and takes the following form
\begin{equation}\label{eq6}
G(x_t,y_t)=\exp\{-V(x_t,y_t)\},\;\;\;\;\;\;    (x_t,y_t)\in (0,\infty)^2,$$ with $$V(x_t,y_t)=2\int_{0}^{1}\max\left(\frac{w}{x_t},\frac{1-w}{y_t}\right)H(dw).
\end{equation}

\nd  $V$ is the exponent measure; for more information, it is preferable to see \cite{Col01,Ha77,Pi81}. Additionally, $H$ is known as a spectral measure on $[0,1]$, which has the best key to control the dependence between the extreme values, and it is a probability measure and obeys the mean constraint
\begin{equation}\label{eq7}
\int_{0}^{1}wH(dw)=\frac{1}{2}.
\end{equation}
If $H$ is absolutely continuous, we define the spectral density as
\begin{equation}\label{eq8}
h(w)=\frac{dH}{dw},\;\;\;\;\;\; w\in [0,1].
\end{equation}

\subsection{Background on Logistic-Normal distribution}\label{sec2.2}
\nd In 1980, Aitchison proposed a technique to approximate a Dirichlet distribution with a Logistic-Normal distribution to minimize their Kullback-Leibler divergence (KL) minimized
\begin{equation}\label{eq9}
K(p,q)=\int_{\Delta_{D}}dir(x|\alpha)\log\left(\frac{dir(x|\alpha)}{h_{\mu,\Sigma}(x)}\right)dx,
\end{equation}
with $dir$ is the density function of the Dirichlet distribution and $h_{\mu,\Sigma}$ is the density function of the Logistic-Normal distribution. This $K(p,q)$ is minimized by
\begin{equation*}
\mu_{i}=\delta(\alpha_{i})-\delta(\alpha_{d+1}),
\end{equation*}
and
\[\sigma_{ij}=
\left\{
\begin{array}{ll}
\epsilon(\alpha_{i})+\epsilon(\alpha_{d+1}),\;\;\;\; i=j\\
\epsilon(\alpha_{d+1}),\;\;\;\; \;\;\;\;\;\;\;\;\;\;\;\; i\neq j
\end{array}
\right.
\]
where $\delta(x)=\Gamma^{'}(x)/\Gamma(x)$, $\epsilon(x)=\delta^{'}(x) ,$ and $\Delta_{d}=\{u\in P^{d}: u_{1}+...+u_{d}<1\}. $\\
\nd The logistic transformation applied to a normal distribution of dimension d produces a distribution on the simplex of dimension d, which can reasonably be called a Logistic-Normal distribution.\\
\nd The Logistic-Normal distribution offers greater flexibility than the Dirichlet distribution, as it can account for correlations between components of probability vectors. This makes it a valuable tool for simplifying statistical analyses of compositional data, as it enables researchers to examine log ratios of data vector components, which are often of interest in comparison to absolute component values.\\
\nd The probability simplex, being a bounded space, poses challenges for applying standard techniques typically used for vectors in $\mathbb {R} ^{n}$. Aitchison highlighted the issue of spurious negative correlations when these methods are directly applied to simplicial vectors. However, transforming compositional data in the probability simplex ${\mathcal {S}}^{D}$ through the inverse of the additive logistic transformation results in real-valued data in ${\mathbb {R}}^{{D-1}}$, where standard techniques can be effectively employed. This method supports the application of the Logistic-Normal distribution, often called the "Gaussian of the simplex."\\

\nd The Logistic-Normal distribution extends the logit-normal distribution to D-dimensional probability vectors by applying a logistic transformation to a multivariate normal distribution.\\
Let $\mathbb{R}^{d}$ represent a d-dimensional real space, $P^{d}$ denote the positive orthant of $\mathbb{R}^{d}$ and $\Delta_{d}$ indicate the d-dimensional positive simplex defined by $\Delta_{d}=\{u\in P^{d}: u_{1}+...+u_{d}<1\}. $\\

\nd For any $d$-dimensional vector $u$ and any real-valued function  $g$, let $g(u)$ signify the $d$-dimensional vector where the $i$th component is given by $g(u_{i}) (i=1,...,d).$ Assuming $v$ follows a multinormal distribution $\mathbf{N}_{d}(\mu,\Sigma)$ over $\mathbb{R}^{d}.$ The exponential transformation from $\mathbb{R}^{d}$ to $P^{d}$, expressed as $w=e^{v},$ alongside its inverse logarithmic transformation $v=\log w,$ serves as the standard method for establishing a corresponding log-normal distribution, where $w$ distributed as $\Lambda_{d}(\mu,\Sigma).$ Similarly, a logistic transformation from  $\mathbb{R}^{d}$ to $\Delta_{d},$ along with its inverse log-ratio transformation
\begin{equation}\label{eq10}
u=e^{v}/\left(1+\sum_{j=1}^{d}e^{v_{j}}\right),\;\;\;\;\; v=\log(u/u_{d+1}),
\end{equation}
where
\begin{equation}\label{eq11}
u_{d+1}=1-\sum_{j=1}^{d}u_{j},
\end{equation}
can be used to define a logistic-normal distribution over $\Delta_{d}$, and we can denote that $u$ follows this distribution $L_{d}(\mu,\Sigma).$ The density function of $L_{d}(\mu,\Sigma)$ is then for $u\in \Delta_{d}$
\begin{equation}\label{eq12}
h_{\mu,\Sigma}(u)=|2\pi\Sigma|^{-\frac{1}{2}}\left(\prod_{j=1}^{d+1}u_{j}\right)^{-1}\exp\left[-\frac{1}{2}\{\log(u/u_{d+1})-\mu\}^{T}\Sigma^{-1}\{\log(u/u_{d+1})-\mu\}\right].
\end{equation}

\nd Even though all positive order moments $\mathbb{E}(u_{j}^{a})\;\;(a>0)$ and the geometric moment
$\exp\{\mathbb{E}(\log u_{j})\}$ exists, but their integral representations cannot be simplified into a straightforward form. However, this is not a significant drawback since practical analysis often focuses more on the ratios $u_{j}/u_{k}$ or their logarithms. Based on normal-logarithmic theory, let $\sigma_{jk}$ represent the $(j,k)$th element of the covariance matrix $\Sigma.$ Additionally, we adopt the convention that $\mu_{d+1}=0$ and $\sigma_{j,d+1}=0$ for $(j=1,...,d+1).$ Consequently, we have that
\begin{equation}\label{eq13}
\mathbb{E}\{\log(u_{j}/u_{k})\}=\mu_{j}-\mu_{k}, \;\;\; \textrm{cov}\{\log(u_{j}/u_{k}),\log(u_{l}/u_{m})\}=\sigma_{jl}+\sigma_{km}-\sigma_{jm}-\sigma_{kl},
\end{equation}
\begin{equation}\label{eq14a}
\mathbb{E}(u_{j}/u_{k})=\exp\{\mu_{j}-\mu_{k}+\frac{1}{2}(\sigma_{jj}-2\sigma_{jk}+\sigma_{kk})\},
\end{equation}
\begin{equation}\label{eq15a}
\textrm{cov}(u_{j}/u_{k},u_{l}/u_{m})=\mathbb{E}(u_{j}/u_{k})\mathbb{E}(u_{l}/u_{m})\{\exp(\sigma_{jl}+\sigma_{km}-\sigma_{jm}-\sigma_{kl})-1\}.
\end{equation}

\subsection{Background on regression manifolds }\label{sec23}

\nd Before introducing our new method of regression of the maxima of the block, we consider $\{(X_{it}, Y_{it})\}_{i=1}^{n}$ a sequence of independent random vectors of unit Fréchet marginal distributions such that $Y_{it}$ is considered response and $X_{it}$ is a covariate. \\Let the componentwise block maxima be $M_{nt}=(M_{x_{t},n},M_{y_t,n})$ with $M_{y_t,n}=\displaystyle\max_{i=1,...,n}\{Y_{it}\}$ and $M_{x_t,n}=\displaystyle\max_{i=1,...,n}\{X_{it}\}.$ \\

\nd Based on the componentwise maxima method introduced by Coles et al. \cite{Col01}, the formulation of the bivariate model is simplified by assuming that both margins adhere to the standard Fréchet distribution. Consequently, the vector of normalized componentwise maxima denoted as $M_{nt}/n$, converges in distribution to a random vector $(X_t, Y_t)$, which is characterized by a bivariate extreme value distribution with the corresponding joint distribution function
\begin{equation}\label{eq15b}
G(x_t,y_t)=\exp\{-V(x_t,y_t)\},\;\;\;\;\; (x_t,y_t)\in (0,\infty)^{2},
\end{equation}
with
\begin{equation}\label{eq16}
V(x_t,y_t)=2\int_{0}^{1}\max\left(\frac{w}{x_{t}},\frac{1-w}{y_t}\right) H(dw).
\end{equation}
The exponent measure is denoted as $V$, while $H$ is a parameter of the bivariate extreme value distribution $G$. This parameter, called the spectral measure, is defined on the interval $[0,1]$ and regulates the dependence between the extreme values. Moreover, $H$ is a probability measure that adheres to the mean constraint

\begin{equation}\label{eq18}
\int_{0}^{1}wH(dw)=\frac{1}{2}.
\end{equation}
If $H$ is absolutely continuous with respect to the Lebesgue measure, then its density can be expressed as the Radon-Nikodym derivative $h=dH/dw,$ where $w$ is defined over the interval $[0,1].$\\

\nd We characterize the regression manifold as the collection of regression lines,
\begin{equation}\label{eq17}
\mathbb{L}=\{\mathbf{L}_{q}:0<q<1\} \; \;\; \mathrm{with}\;\;\;   \mathbf{L}_{q}=\{y_{q|x_t}:x_t \in (0,\infty)\},
\end{equation}

\nd where
\begin{equation}\label{eq19}
y_{q|x_t}=\inf\{y_t>0, G_{Y_t|X_t}(y_t|x_t)\geq q\},
\end{equation}

\nd is a conditional quantile of bivariate extreme value distribution, with $q \in (0,1)$ and $x_t\in (0,\infty),$ and $G_{Y_t|X_t}(y_t|x_t)=\mathbb{P}(Y_t\leq y_t|X_t=x_t)$ is a conditional bivariate extreme value distribution function as defined by
\begin{equation}\label{eq20}
G_{Y_t|X_t}(y_t|x_t)=2\exp \left\{-2\int_{0}^{1}\max\left(\frac{w}{x_t},\frac{1-w}{y_t}\right)h(w)dw+x_t^{-1}\right\}\int_{w(x_t,y_t)}^{1}wh(w)dw.
\end{equation}

\begin{figure}[h]
\centering
\includegraphics[width=1.1\textwidth]{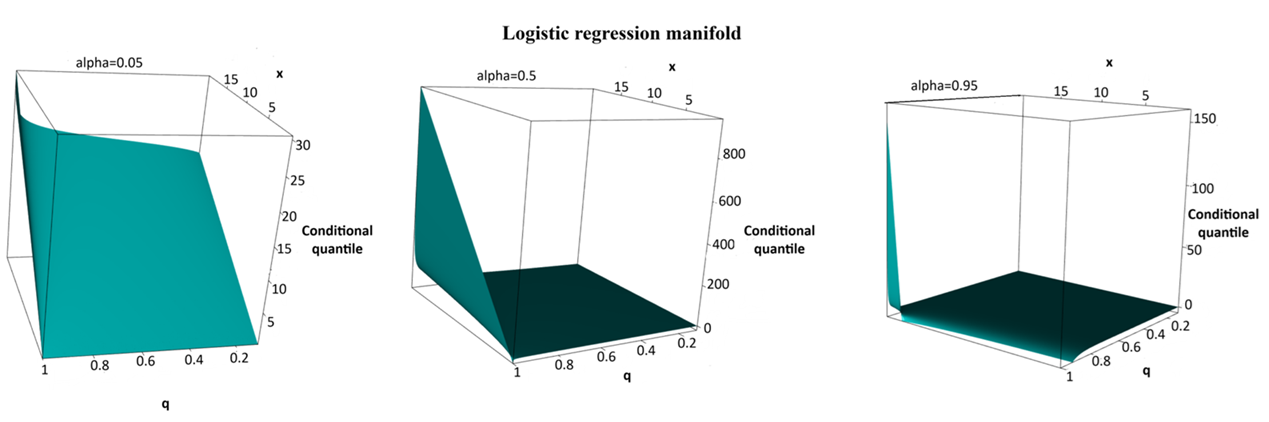}
\caption{Regression manifold for bivariate logistic. }\label{1:f1}
\end{figure}

\nd We will now examine certain parametric instances of regression manifolds as recently defined.
\begin{exm}[Non-stationary logistic model]
An instance of the non-stationary Logistic regression manifold is shown in Fig. \ref{1:f1}. This follows from the non-stationary Logistic bivariate extreme value distribution function given by \begin{equation}\label{eq14b}
G(x_t,y_t)=\exp\{-(x_t^{-1/\alpha}+y_t^{-1/\alpha})^{\alpha}\}, \; \;\;  \; \;\;  x_t,y_t>0,
\end{equation}
where $\alpha \in (0,1]$ represents the strength of dependence between extremes. As $\alpha$ approaches 0, the dependence becomes stronger, with $\alpha\rightarrow 0$ indicating perfect dependence. The conditional distribution of $Y_t$ given $X_t$ is
\begin{equation}\label{eq21}
G_{Y_t|X_t}(y_t|x_t)=G(x_t, y_t)(x_t^{-1/\alpha}+y_t^{-1/\alpha})^{\alpha-1}x_t^{1-1/\alpha}\exp(1/x_t), \; \;\;  \; \;\; x_t,y_t>0.
\end{equation}

\nd For regression manifold representation, we use the approximation of the exact logistic regression manifolds defined by Carvalho et al. \cite{Ca22} as follows for $q \in (0,1)$ and $x_t>>1$
\begin{equation}\label{eq22}
\tilde{y}_{q|x_t}=\frac{\alpha}{1-\alpha}\{q^{1/(\alpha-1)}-1\}^{-\alpha-1}\{q^{\alpha/(1-\alpha)}-1\}q^{1/(\alpha-1)}+\{q^{-1/(1-\alpha)}-1\}^{-\alpha}x_t.
\end{equation}

\end{exm}

\begin{figure}[h]
\centering
\includegraphics[width=7.5in,height=2.5in]{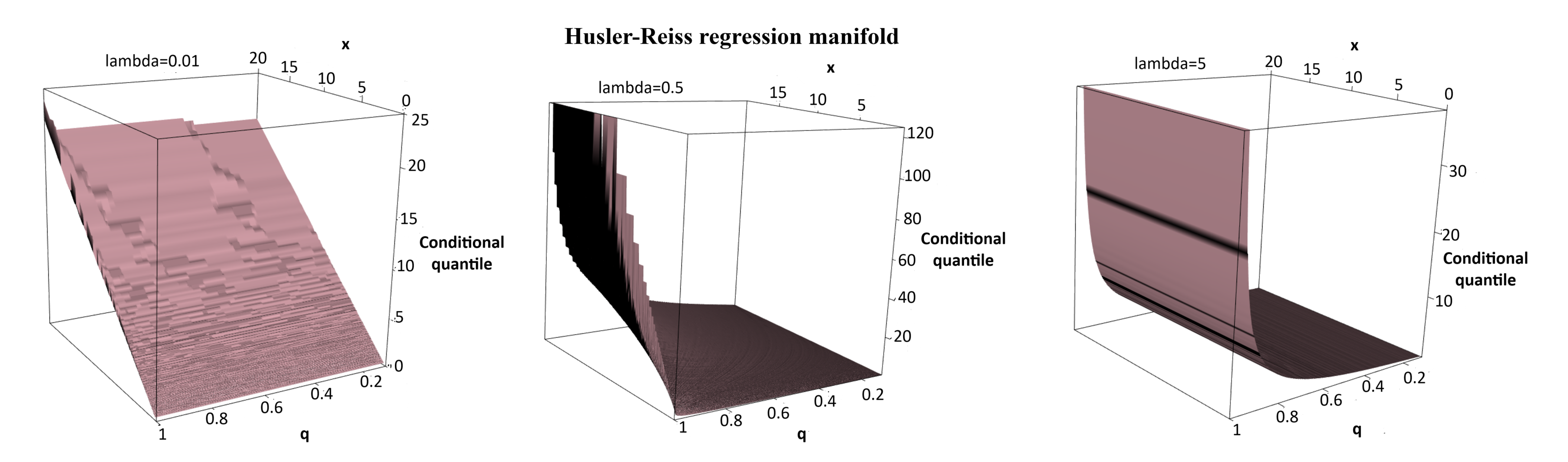}
\caption{Regression manifold for bivariate Husler-Reiss. }\label{1:f2}
\end{figure}

\begin{figure}[h]
\centering
\includegraphics[width=7in,height=2.5in]{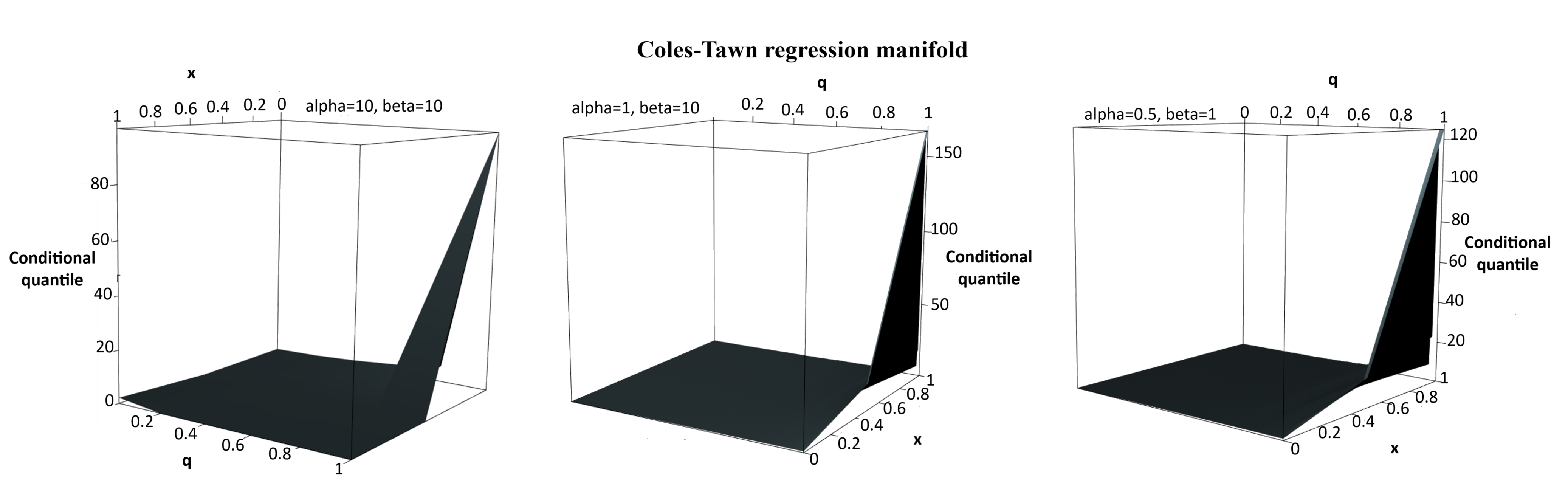}
\caption{Regression manifold for bivariate Coles-Tawn.}\label{1:f3}
\end{figure}

\begin{exm}[Non-stationary Husler-Reiss model]
An instance of the non-stationary Husler-Reiss regression manifold is shown in Fig. \ref{1:f2}. It arises from the non-stationary Husler-Reiss bivariate extreme value distribution function, which has the following form
\begin{equation}\label{eq23}
G(x_t,y_t)=\exp\left\{-x_t^{-1}\Phi\left(\lambda+\frac{1}{2\lambda}\log\frac{y_t}{x_t}\right)-y_t^{-1}\Phi\left(\lambda+\frac{1}{2\lambda}\log\frac{x_t}{y_t}\right)       \right\}, \; \;\; \; \;\; x_t,y_t>0,
\end{equation}
where $\Phi$ denotes the standard normal distribution function and $\lambda \in (0,\infty]$ is a parameter that modulates the dependency between extremes: as $\lambda\rightarrow 0$, the dependence becomes perfect, while in the limit as $\lambda\rightarrow\infty$, the dependency reaches complete independence. The collection of regression lines  $\mathbf{L}_{q}$ for this model lacks explicit forms and is derived using (\ref{eq19}) with
\begin{equation}\label{eq24}
\small
G_{Y_t|X_t}(y_t|x_t)=\left[\Phi\left(\lambda+\frac{1}{2\lambda}\log\frac{y_t}{x_t}\right)+\frac{1}{2\lambda}\phi\left(\lambda+\frac{1}{2\lambda}\log\frac{y_t}{x_t}\right)-\frac{x_t y_t^{-1}}{2\lambda}\phi\left(\lambda+\frac{1}{2\lambda}\log\frac{x_t}{y_t}\right)\right]\times G(x_t,y_t)\exp(1/x_t),
\end{equation}
for $x_t,y_t>0$ where $\phi$ is the standard Normal density function.
\end{exm}
\begin{exm}[Non-stationary Coles-Tawn model]
An instance of the non-stationary Coles-Tawn regression manifold is shown in Fig. \ref{1:f3}. This idea is derived from the non-stationary Coles-Tawn bivariate extreme value distribution function, which is defined as follows
\begin{equation}\label{eq5}
G(x_t,y_t)=\exp[-x_t^{-1}\{1-Be(q;\alpha+1,\beta)\}-y_t^{-1}Be(q;\alpha,\beta+1)], \; \;\;  \; \;\; x_t,y_t>0,
\end{equation}
where $Be(q; a,b)$ represents the cumulative distribution function of a $Beta$ distribution with parameters $a,b>0$. In this context,
$q$ is defined as $q=\alpha y_t^{-1}/(\alpha y_t^{-1}+\beta x_t^{-1})$ with $\alpha, \beta>0$ serving as parameters that control the dependence between extremes. The scenario where $\alpha=\beta=0$  indicates complete independence, while the case of $\alpha=\beta\rightarrow\infty$ signifies perfect dependence. For fixed values of $\alpha(\beta)$, an increase in $\beta(\alpha)$  leads to a stronger dependence. The family of regression lines $\mathbf{L}_{q}$ for this model does not have a straightforward representation and is determined through calculations using (\ref{eq19}) for $x_t, y_t > 0$, with

\small
$$G_{Y_t|X_t}(y_t | x_t)= \left[ 1-Be(q;\alpha+1,\beta) +\frac{(\alpha+1)\beta}{\gamma}be(q;\alpha+2,\beta+1)-
\frac{x_t}{y_t}\frac{\alpha(\beta+1)}{\gamma}be(q;\alpha+1,\beta+2)\right]$$
\begin{equation}\label{eq27}
G(x_t,y_t)\exp(1/x_t).
\end{equation}

\end{exm}
\subsection{Induced prior on the space of regression manifolds via Logistic-Normal distribution}\label{sec4}

\nd In this section, we explore the process of understanding regression manifolds through data analysis. In this context, de Carvalho et al. \cite{Ca22} introduced a prior related to the space of spectral measures put forth by Hanson et al. \cite{Han17}. The model is a Bernstein polynomial whose expression contains the Dirichlet distribution. In addition, Aitchison \cite{ait80} proposed the famous relationship between the Dirichlet distribution and Logistic-Normal distribution and proved the flexibility of this last. So, we introduce our new model.\\

\nd Let's $\{(X_{it}, Y_{it})\}_{i=1}^{n}$ be a non-stationary sequence of independent random vectors with unit Fréchet marginal distributions. We estimate our parameters (scale, location, and shape) with the transformed-stationary approach proposed by Mentaschi et al. \cite{Men16}.\\
Define $R_{it}=Y_{it}+X_{{it}}$ and $W_{it}=(X_{it},Y_{it})/R_{it},$ as the pseudo-angular decomposition of the observations. De Haan and Resnick \cite{Ha77} demonstrated that the convergence of normalized componentwise maxima to $G$ is equivalent to a specific weak convergence of measures and generalized the result for the time series
$$\mathbb{P}(W_t\in .|R_t>u_t)\xrightarrow{d} H(.),\;\;\; \; \;\; \textrm{as}\; u_t\rightarrow \infty.$$
This implies that when the radius $R$  is large enough, the pseudo-angles $W$ become almost independent of $R$ and resemble a distribution linked to the spectral measure $H$. Therefore, to learn about $\mathbf{L}_{q}$, we initially need to investigate $H$ by analyzing $k_t=|\{W_{it}: R_{it}>u_t, i=1,...,n\}|$ which counts the number of exceedances above a high threshold $u_t$, where $k_t=o(n).$ We will outline the methodology for this analysis next.\\

\nd In this paper, we model the spectral density $h$ via the spectral measure via the Logistic-Normal distribution for obtaining an more admirable and important estimation that can thus be considered as the "simplex Gaussian". Find the right simplex to define our fabulous spectral density $ h. $

\nd Firstly, we know the Logistic-Normal distribution is defined on the d-dimensional positive simplex defined by $\Delta_{d}=\{w\in P^{d}: w_{1}+...+w_{d}<1\}. $ We add at this simplex $w_{d+1}=1-\sum_{i=1}^{d}w_i.$ We obtain $\Delta_{d+1}=\{w\in P^{d}: w_{1}+...+w_{d}+w_{d+1}=1\}. $ So, we consider the simplex $S_{d}=\{(w_{1},...,w_{d})\in [0,1]^{d}, \sum_{i=1}^{d} w_{i}=1\}\subset \mathbb{R}^{d}.$ We can work for that because the $\{w_{i}\}_{i=1}^{d}$ and $w_{d+1}$ are linear. So, in our case, we take $d=1.$\\

\nd For modeling our spectral density $h$ via the Logistic-Normal distribution we must first estimate $\mu$ and $\sigma$ the Logistic-Normal parameters'. Therefore, the next lemma presents a good result for estimating the parameter $\mu.$
\begin{lem}
We consider $h_{\mu,\sigma}(u)=\frac{\exp\left(-\frac{\log^2\left(\frac{u}{1-u}-\mu\right)}{2\sigma^2}\right)}{\sigma\sqrt{2\pi}u(1-u)}$  then $\int_0^1 uh_{\mu,\sigma}(u)dw=1/2\Leftrightarrow \mu=0.$
\end{lem}
\begin{proof}[Proof]
We consider $\mathbb{E}(u)=\int_{0}^{1}th_{\mu,\sigma}(t)=\frac{1}{2}$ and $\mathcal{Z}\sim\mathcal{N}(0,1).$\\

\nd  So $$\mathbb{E}\left(\frac{e^{\sigma\mathcal{Z}+\mu}}{1+e^{\sigma\mathcal{Z}+\mu}}\right)=\frac{1}{2}=\mathbb{E}\left(\frac{e^{\sigma\mathcal{Z}}}{1+e^{\sigma\mathcal{Z}}}\right).$$
We know $\mathcal{\mathcal{Z}}\overset{\mathcal{L}}=\mathcal{-Z}.$
Then $$\mathbb{E}\left(\frac{e^{-\sigma\mathcal{Z}+\mu}}{1+e^{-\sigma\mathcal{Z}+\mu}}\right)=\mathbb{E}\left(\frac{1}{1+e^{\sigma\mathcal{Z}}}\right)$$ Thus $$\mathbb{E}\left(\frac{e^{\sigma\mathcal{Z}+\mu}}{1+e^{\sigma\mathcal{Z}+\mu}}\right)=\mathbb{E}\left(\frac{1}{1+e^{\sigma\mathcal{Z}}}\right)$$
$$\Leftrightarrow \mathbb{E}\left(\frac{1}{1+e^{\sigma\mathcal{Z}+\mu}}\right)-\mathbb{E}\left(\frac{1}{1+e^{\sigma\mathcal{Z}}}\right)=0$$
$$\Leftrightarrow \mathbb{E}\left(\frac{e^{\sigma\mathcal{Z}}(1-e^\mu)}{(1+e^{\sigma\mathcal{Z}+\mu})(1+e^{\sigma\mathcal{Z})}}\right)=0$$
$$\Leftrightarrow (1-e^\mu)\mathbb{E}\left(\frac{e^{\sigma\mathcal{Z}}}{(1+e^{\sigma\mathcal{Z}+\mu})(1+e^{\sigma\mathcal{Z}})}\right)=0$$
$$\Leftrightarrow \mu=0$$ because $\frac{e^{\sigma\mathcal{Z}}}{(1+e^{\sigma\mathcal{Z}+\mu})(1+e^{\sigma\mathcal{Z}})}>0 $ almost surely.\\

\nd  If $\mu=0$, we solve the integral $\int_{0}^{1}wh_{0,\sigma}(w)dw.$ After all calculation, we find that the integral is $1/2$ hence the result.
\end{proof}

\nd Now, we model the spectral measure density $h$ by the expression
\begin{equation}\label{eq29}
h(w)=h_{0,\sigma}(w),\;\;\; \; \;\; w\in [0,1].
\end{equation}

\nd Note the parameters $\sigma$ is estimated by obeying at the moment constraint $\int_{0}^{1}wH(dw)=\frac{1}{2}.$
We use the logistic transformation $w=e^{v}/(1+e^{v})\;\;\; \textrm{and}\;\;\; v=\log(w/1-w)$ with $v$ follows a normal distribution $\mathbf{N}(0,\sigma).$ \\

\nd For estimating the parameter $\sigma$, we use the maximum likelihood estimation (MLE) with
\tiny$$\log L(x_t,y_t)=\sum_{i=1}^{n}\log\left[\frac{2}{x_{it}^2y_{it}^2\sigma^2\pi}\int_{w(x_t,y_t)}^{1}\frac{\exp\left(-\frac{\log^2\left(\frac{w}{1-w}\right)}{2\sigma^2}\right)}{1-w}dw\int_{w(y_t,x_t)}^{1}\frac{\exp\left(-\frac{\log^2\left(\frac{w}{1-w}\right)}{2\sigma^2}\right)}{1-w}dw+\sqrt{\frac{2}{\pi}}\times \frac{1}{\sigma(x_{it}+y_{it})x_{it}y_{it}}\exp\left(-\frac{\log^2\left(\frac{x_{it}}{y_{it}}\right)}{2\sigma^2}\right)\right]\\$$
\begin{equation}
-\frac{2}{x_i\sigma \sqrt{2\pi}}\int_{w(x_t,y_t)}^{1}\frac{\exp\left(-\frac{\log^2\left(\frac{w}{1-w}\right)}{2\sigma^2}\right)}{1-w}dw-\frac{2}{y_{it}\sigma \sqrt{2\pi}}\int_{w(y_t,x_t)}^{1}\frac{\exp\left(-\frac{\log^2\left(\frac{w}{1-w}\right)}{2\sigma^2}\right)}{1-w}dw,
\end{equation}\normalsize\label{eq30}

\nd and we maximize the function to obtain a new estimator $\hat{\sigma}.$\\

\nd Now, to incorporate a prior into the space of regression manifolds, we substitute the spectral density \begin{equation}\label{31}
h(w)=h_{0,\hat{\sigma}}(w),\;\;\;  \; \;\; w\in [0,1],
\end{equation}into (\ref{eq20})
with $w(x_t,y_t)=x_t/(x_t+y_t)$ for $x_t,y_t>0.$   \\

\nd This prior induces a prior on the space of regression lines $\mathbf{L}_{q}=\{y_{q|x_t}: x_t\in (0,\infty)\},$ where $y_{q|x_t}$ is a solution to equation, $G_{Y_t|X_t}(y_t|x_t)=q$, for $q \in (0,1)$, where\begin{equation}\label{eq32}
G_{Y_t|X_t}(y_t|x_t)=2\exp \left\{-2\int_{0}^{1}\max\left(\frac{w}{x_t},\frac{1-w}{y_t}\right)h_{0,\hat{\sigma}}(w)dw+x_t^{-1}\right\}\int_{w(x_t,y_t)}^{1}wh_{0,\hat{\sigma}}(w)dw,
\end{equation}
with $w(x_t,y_t)=x_t/(x_t+y_t)$ for $x_t,y_t>0.$\\

\section{Simulation study}\label{sec3}
\subsection{Preliminary experiments}\label{sec31}
\begin{figure}[!h]
\centering
\includegraphics[width=6in,height=6in]{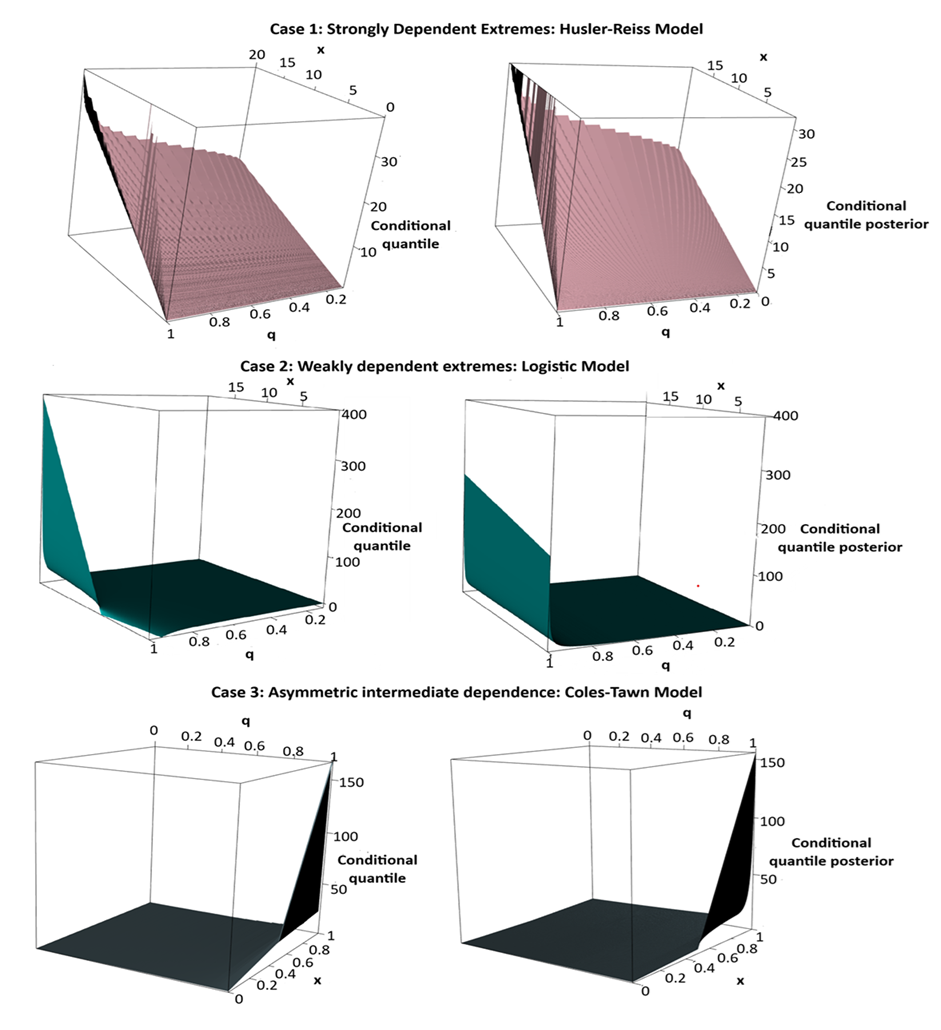}
\caption{The true regression manifold $\mathbb{L}$ (shown on the left) along with its posterior mean estimate (shown on the right) is presented for the Husler–Reiss model (top), the Logistic model (middle), and the Coles–Tawn bivariate extreme value models (bottom), as derived using the methods discussed in Sect. \ref{sec4}. }\label{1:f4}
\end{figure}

\nd As follows de Carvalho et al. \cite{Ca22}, we examine the finite sample performance of the proposed methods across three data generating scenarios outlined in Section \ref{sec23}, specifically in Examples 1–3. To accomplish this, we simulate the data in the following manner:\\
\begin{itemize}
\item Case 1: strongly dependent extremes: Husler-Reiss model with $\lambda=0.1$;
\item Case 2: weakly dependent extremes: Logistic model with $\alpha= 0.9$;
\item Case 3: asymmetric intermediate dependence: Coles–Tawn model with $\alpha=0.5$, $\beta=100.$
\end{itemize}
To demonstrate the comparison between the resulting estimates and the true regression lines in a single experiment for each case, we generate $n=2000$ non-stationary samples $\{(\mathcal{X}_{it}, \mathcal{Y}_{it})\}_{i=1}^n$ with $t\in [i-\frac{n}{2},i+\frac{n}{2}]$. After that, we use the transformed-stationary approach developed by Menstachi et al. \cite{Men16}: transforming $\{(\mathcal{X}_{it}, \mathcal{Y}_{it})\}_{i=1}^n$ to a stationary series $\{(\tilde{X}_{it}, \tilde{Y}_{it})\}_{i=1}^n$, fitting a GEV model from stationary data, fitting time-varying GEV parameters (see the equations \eqref{4}, \eqref{5}, \eqref{6}) and finally transforming the stationary data to the non-stationary data $\{(X_{it}, Y_{it})\}_{i=1}^n$ by fitting a new GEV model via the new time-varying parameters. For the analysis, we utilize observations for which $\hat{X}_{it}+ \hat{Y}_{it}> u_t$, whith $u_t$ being the $98\%$ quantile. The raw data is transformed into unit Fréchet margins using the transformation $(\hat{X}_{it},\hat{Y}_{it})=(-1/\log\{
\hat{F}_X(X_{it})\},-1/\log\{\hat{F}_Y(Y_{it})\})$, where $\hat{F_X}$ and $\hat{F_Y}$ represent the empirical distribution functions, normalized by $n+1$ to prevent division by zero. We plot each case's true and estimated regression manifolds; see Fig \ref {1:f4}. For estimated regression manifolds, we are modeling the spectral measure by Logistic-Normal distribution with $\mu=0$ and maximizing the loglikelihood function to obtain the estimator  $\hat{\sigma}$. Fig. \ref{1:f4} illustrates that, for these one-shot experiments, the proposed methods accurately capture the shape of $\mathbb{L}$ in all three cases. However, as expected, the fits are less precise when $q$ is close to 0 or 1.

\subsection{Real data illustrations}\label{sec6}
\subsubsection{Data analysis}\label{sec6.1}
\nd We now implement the proposed method to analyze climatic extremes. In particular, we generate data using the following approach: We work with $n=50$ componentwise maxima for precipitation and temperature is denoted below as $\{(\mathcal{X}_{it}, \mathcal{Y}_{it})\}_{i=1}^{n}$ with $t \in [i-\frac{n}{2},i+\frac{n}{2}]$. Similarly to Sect.\ref{sec31}, we applied the transformed-stationary approach seen in Sect.\ref{sec2.1} proposed by Mentaschi et al. \cite{Men16}, which consists of three steps: first, transforming the non-stationary time series $\{(\mathcal{X}_{it}, \mathcal{Y}_{it})\}_{i=1}^n$
into a stationary series $\{(\tilde{X}_{it}, \tilde{Y}_{it})\}_{i=1}^n$. Next, performing a stationary extreme value analysis $\{(X_{it}, Y_{it})\}_{i=1}^n$ and finally, reverting the resulting extreme value distribution to a time-dependent form. \\
First, we transform the negative log returns into unit Fréchet margins using the transformation
$(\hat{X_{it}},\hat{Y_{it}})=(-1/\log\{\hat{F_{X}}(X_{it})\},-1/\log\{\hat{F_{Y}}(Y_{it})\}),$ where $\hat{F_{X}}$ and $\hat{F_{Y}}$ represent the empirical distribution functions of the negative log returns for precipitation $(Y_t)$ and temperature $(X_t)$, respectively. These functions are normalized by $n+1$ instead of $n$ to avoid division by zero.\\

\begin{figure}[!h]
\centering
\includegraphics[width=1\textwidth]{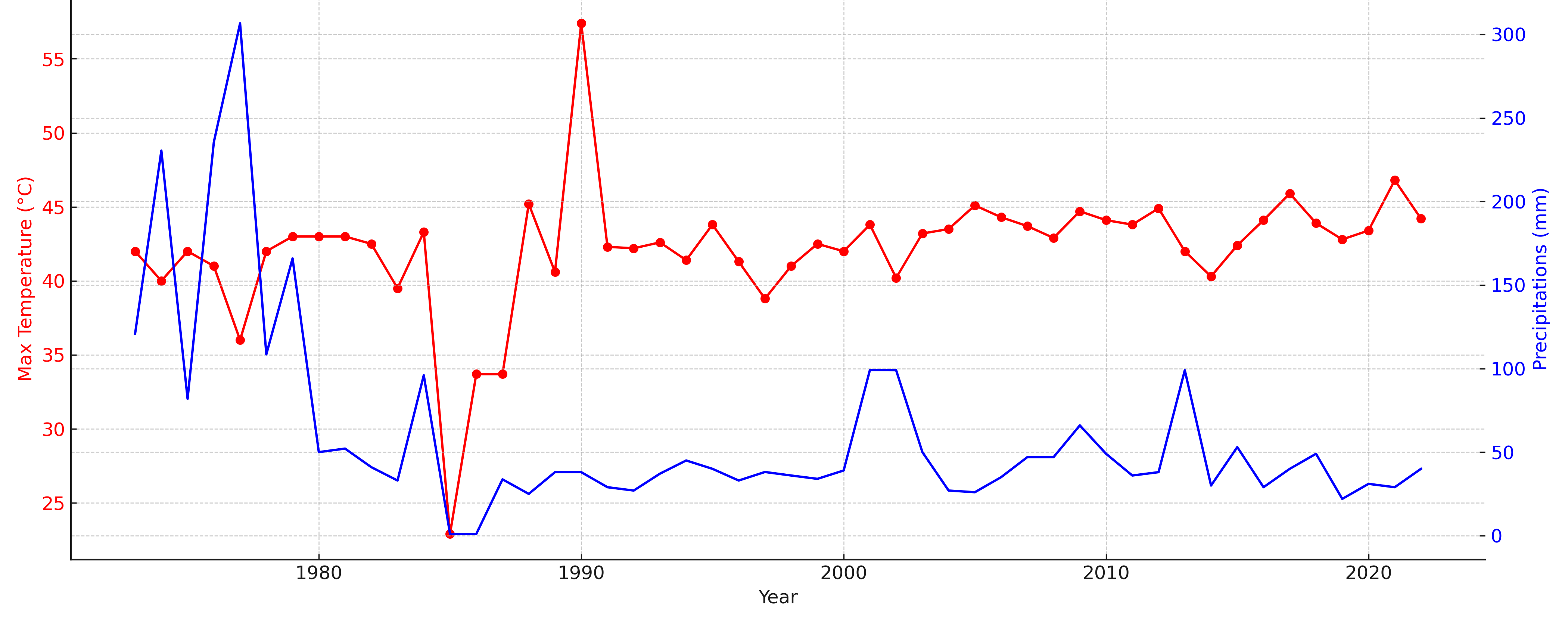}
\caption{True data of yearly Max temperature (red) and precipitations (blue).  }\label{1:f6}
\end{figure}

\begin{figure}[!h]
\centering
\includegraphics[width=6in,height=3.5in]{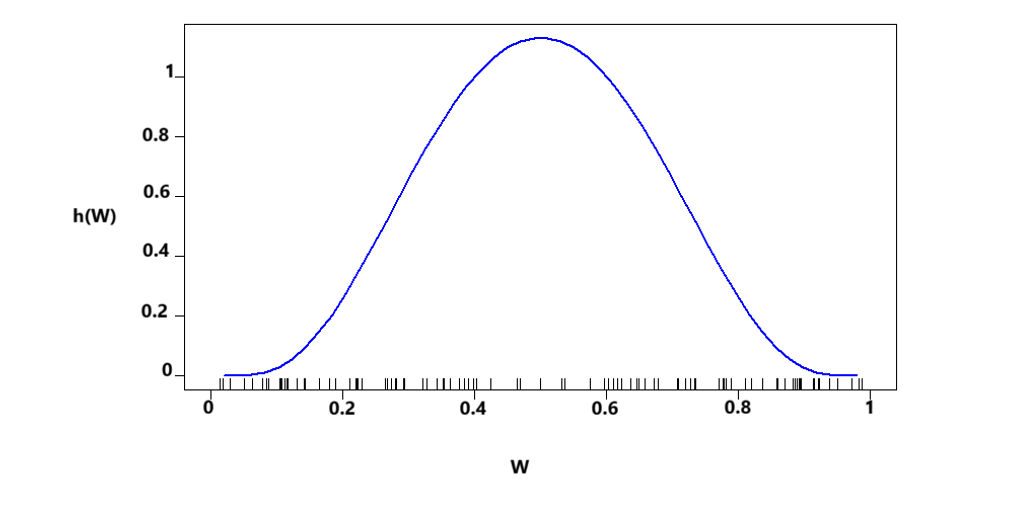}
\caption{Spectral density estimate with $95\%$ credible band along with a rug of pseudo-angles.}\label{1:f8}
\end{figure}

\nd The real data of maxima of precipitation and temperature have been taken between 1973-2022 from the local Meknes-Morocco. The red one represents the real temperature data, but the blue one represents the real precipitation data; see Fig. \ref {1:f6}. We present the spectral density estimate in Fig. \ref {1:f8}. The plot shows that most of the observed pseudo angles are concentrated in the middle of the interval $(0, 1).$ The distribution resembles a right-skewed bell-shaped curve, indicating an asymmetric intermediate dependence between the extremal losses of the temperature and precipitation composite indices. This asymmetry suggests that when both indices experience extreme conditions, the losses on the precipitation tend to be more severe than those on the temperature.

\subsubsection{Modeling regression manifolds }\label{sect.6.2}
\begin{figure}[!h]
\centering
\includegraphics[width=7in,height=3.5in]{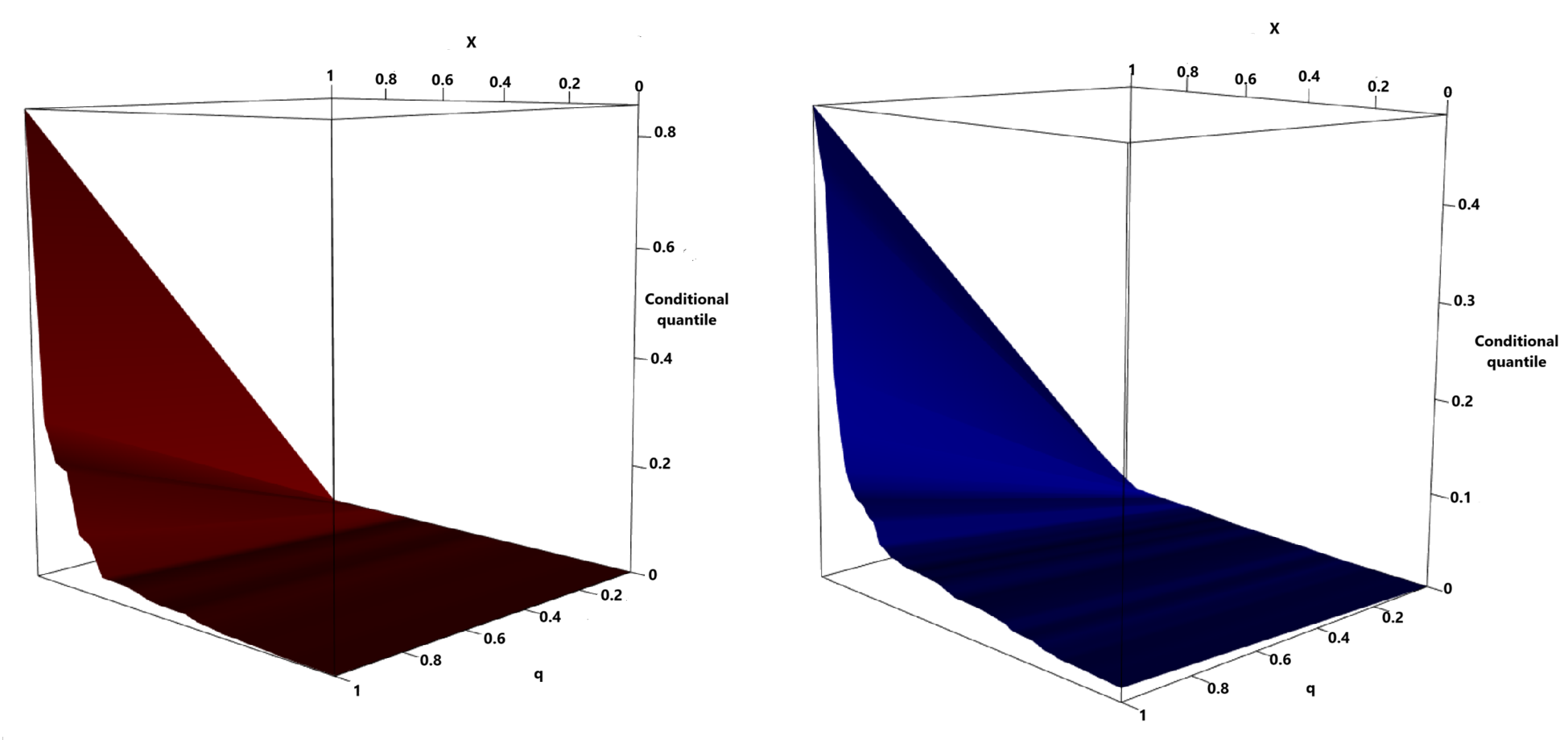}
\caption{Estimated regression manifold (left) and Carvalho et al.'s (right) on the original scale for precipitations given temperature. }\label{1:f7}
\end{figure}

\nd In this section, we will show how the regression manifold can be used to model the dependence between extremes on the precipitation given an extreme on the temperature. We illustrate the posterior mean regression manifold on the original scale for precipitation given temperature; see Fig. \ref {1:f7}. We contrasted our model with the one put forth by Carvalho et al. \cite{Ca22} to show how effective it is. For that, we use a typical componentwise adaptive Markov Chain Monte Carlo (MCMC) \cite{har05} with a Dirichlet prior, Dirichlet($10^{-4}\mathbf{1_k}$) defined on a generalized logit transformation of weights $\pi_\alpha$ to learn about regression lines from data. Every MCMC chain has a burn-in duration of 4,000 and a length of 10,000.\\
\begin{table}[!ht]
\centering

\begin{tabular}{lll}
\hline

Goodness model & New model& Carvalho et al. \cite{Ca22} \\\hline
AIC & $280.7239$ &$354.8581$ \\ \hline
BIC & $286.4599$ & $362.5062$  \\ \hline

\end{tabular}
\caption{The comparison of the estimated regression manifolds for the new model and Carvalho et al. \cite{Ca22}  using AIC and BIC criteria. }
\label{tab3}
\end{table}

\begin{table}[!ht]
\centering

\begin{tabular}{llll}
\hline
Precipitations & Temperature&  & ~ \\\hline
~ & $1\%$ & $2\%$ & $3\%$ \\ \hline
$75\%$ &$2,006487\times 10^{-8}$  & $9,990657\times 10^{-8}$ & $3,108382\times 10^{-7}$\\\hline
$90\%$ & $2,006706\times 10^{-8}$ & $9,992477\times 10^{-8}$ & $3,109213\times 10^{-7}$ \\ \hline
$95\%$ & $2,006706\times 10^{-8}$ & $9,99248\times 10^{-8}$ & $3,109214\times 10^{-7}$ \\ \hline
\end{tabular}
\caption{The predicted $75\%$, $90\%$ and $95\%$ quantiles of losses related to precipitation are assessed for $1\%$, $2\%$ and $3\%$ weekly maximum losses in temperature, with $95\%$ credible intervals indicated in brackets. Negative log returns are employed as a proxy for these losses.}
\label{tab2}
\end{table}

\nd As can be seen from figure \ref{1:f7}, our new model with non-stationary extremes shows the same result when the extremes are stationary. In addition, it is identical to the estimated regression manifolds model proposed by Carvalho et al. \cite{Ca22}. Moreover, as shown in Fig. \ref{1:f7}, the regression manifold exhibits significant non-linearity. In the middle graph, the regression lines differ markedly from those associated with independence and are generally closer to the identity line. We evaluated our new model against the one proposed by Carvalho et al. \cite{Ca22} using the AIC and BIC criteria, which indicated that our model is more effective, as evidenced by its lowest AIC and BIC values.
Table \ref{tab2} presents the predicted $75\%$, $90\%$, and $95\%$ quantiles of losses on the precipitation, evaluated against $1\%$, $2\%$, and $3\%$ maximum weekly losses on the temperature. This analysis is derived from the regression manifold and can be interpreted as follows:\\
\nd Firstly, from a qualitative perspective, Table \ref{tab2} suggests that when the temperature experiences a significant decline, the precipitation is likely to experience a comparable decrease. Secondly, and more importantly, from a climatic analysis standpoint, the table provides quantitative insights. For instance, it indicates that in instances of a $1\%$ weekly maximum loss on the temperature, there is only a $5\%$ chance that the loss on the precipitation will exceed $0.000002\%$. Similarly, the table reveals that in cases with a $3\%$ weekly loss on the temperature, the likelihood of experiencing a loss on the precipitation greater than $0.00003\%$ is only $5\%$. This shows that a decrease in temperature does not lead to a decrease in precipitation, as evidenced by the low percentages obtained. On the contrary, precipitation tends to be low at high temperatures, which explains, according to our statistical study, a shallower slope of extreme precipitation with the increase in extreme temperatures. Thus, we can consider this study a warning for humanity to combat all actions contributing to the greenhouse effects that threaten our planet. As anticipated, the results are equivalent to those presented here. So, our model is better, especially when we use logistic-normal distribution, which can be seen as the "Gaussian of the simplex".

\section*{Conclusion}
\nd We introduce a regression model designed for cases in which both the response variable and the covariate exhibit extreme behavior. The foundation of this model is based on the observation that the limiting behavior of the vector formed by appropriately standardized componentwise maxima follows a bivariate extreme value distribution. The construction of the model parallels the approach used in quantile regression, where we evaluate how the conditional quantile of the response variable responds to changes in the covariate, taking into consideration the asymptotic results. A key focus of this proposed framework is the regression manifold, which comprises a set of regression lines that adhere to bivariate extreme value theory principles. Our model uses regression techniques to analyze data where the response and the covariate are non-stationary extremes. To analyze the non-stationary extremes, we use the T-S methodology developed by Mentaschi et al. \cite{Men16}, which consists of three steps: first, transforming the non-stationary time series $y_t$
into a stationary series $x_t$; next, a stationary extreme value analysis (EVA) is performed; and finally, the resulting extreme value distribution is reverted to a time-dependent form. The model is based on the idea that the behavior of properly standardized maximum values follows a bivariate extreme value distribution. Similar to quantile regression, the model evaluates how changes in the covariate affect the conditional quantile of the response while considering this asymptotic result. The model focuses on the regression manifold, which consists of regression lines that adhere to bivariate extreme value theory. A logistic-normal distribution is employed as a prior in the context of spectral densities to extract insights from the data regarding the model, with numerical analyses demonstrating its versatility. We use a prior on the space of spectral densities to learn about the model from data, and numerical studies demonstrate its flexibility. In this paper, we worked with statistical models for non-stationary extremes \cite{Col01}, an alternative to the methods proposed by de Carvalho et al. \cite{Ca22}. Numerical studies proved the goodness of our model with AIC and BIC criteria. These models can indeed assess the impact of covariates on an extreme-valued response by linking the parameters of the Generalized Extreme Value (GEV) distribution to a covariate. However, since these models are based on univariate extreme value theory, they are not specifically designed for scenarios where another variable is also extreme. They overlook the information arising from the dependence structure between extremes. Similarly, extremal quantile regression methods \cite{Ch05}, like the non-stationary extreme models, do not account for conditioning on another extreme variable, as they also neglect the dependence structure among the extremes. To better visualize our well-known method, we have worked on climate extremes to guard against the dangers and risks that threaten us in the future due to climate change.

\nd A promising direction for future research is the introduction of a new regression model for non-stationary extremes using advanced regression techniques based on neural networks and deep learning. These methods provide greater flexibility and improved generalization capabilities for capturing complex relationships among extreme variables. By incorporating neural network architectures, it becomes possible to develop models that automatically learn underlying data structures, enhancing the accuracy of extreme event predictions. Special attention could be given to convolutional and recurrent neural networks to account for spatial and temporal trends, thus improving adaptability to dynamic climate variations.

\section*{Data Availability}

\nd The datasets analyzed in this study can be accessed from Meteociel and Infoclimat free websites: the period 1973-1979 of these data are downloaded from \url{https://www.meteociel.fr/climatologie/obs_villes.php?code2=60150&mois=12&annee=1973}, but the other period 1980-2022 is downloaded from \url{https://www.infoclimat.fr/climatologie/annee/1980/meknes/valeurs/60150.html}, which has authorized us to use them.\\

\section*{Conflicts of Interest}

\nd The authors declare that they have no conflicts of Interest.
\bibliographystyle{plainnat}

\end{document}